\documentclass[sigconf]{acmart}
\AtBeginDocument{%
  }

\begin{document}

\title[AI Fortune Teller]{AI Fortune Teller: Juxtaposing Shaman and AI to Reveal Human Agency in the Age of AI}

\author{Soonho Kwon}
\email{soonho@gatech.edu}
\orcid{0000-0002-2783-6364}
\affiliation{%
  \institution{Georgia Institute of Technology}
  \city{Atlanta}
  \state{GA}
  \country{USA}}

\author{Dong Whi Yoo}
\email{dy22@iu.edu}
\orcid{-}
\affiliation{%
  \institution{Indiana University Indianapolis}
  \city{Indianapolis}
  \state{IN}
  \country{USA}}

\author{Younah Kang}
 \email{yakang@yonsei.ac.kr}
 \affiliation{%
   \institution{Yonsei University}
   \city{Seoul}
   \country{Republic of Korea}}

\renewcommand{\shortauthors}{Kwon et al.}

\begin{abstract}
This video piece showcases participants interacting with a career counseling AI agent, unaware that the responses were actually derived from the fortunetelling of a \textit{mudang} (a Korean traditional shaman). Our work captures this deception and documents participants’ reactions, showcasing shifts in their initial perceptions of the agent’s advice following the reveal. Notably, even after learning that the advice came from a \textit{mudang} rather than an AI, participants did not change their initial attitudes toward the advice they received. This raises questions about the perceived importance of AI’s explainability and accuracy. By juxtaposing scientific and pre-scientific approaches, we aim to provoke discussions on human agency in the age of AI. We argue that, regardless of AI’s advancements, we continue to navigate life in fundamentally human ways—wonderfully messy and uncertain.

\textbf{   }

\noindent\textcolor{blue}{\textbf{Watch AI Fortune Teller: \href{https://www.youtube.com/watch?v=Y21DGAZQWMU}{ENG} | \href{https://www.youtube.com/watch?v=qv34Cx4Snwc}{KOR}}}

\textbf{   }

\noindent\textit{Disclaimer: This document is an unofficial commentary on AI Fortune-Teller by its creators. While the work was introduced and received an Honorary Mention at Prix Ars Electronica 2024, this document is not an officially published or affiliated record of the festival.}

\end{abstract}




\maketitle

\section{Introduction}

The way people interact with artificial intelligence (AI) agents today, particularly those that leverage predictive algorithms, bears a striking resemblance to how they engage with traditional fortunetelling \cite{cho2025shamain}. People turn to both predictive algorithms and fortunetelling to make sense of and seek guidance for an uncertain future. In doing so, they both utilize "data" of the present, and the reasoning or mechanism behind their recommendations often remains opaque to those who receive them.

Predictive AI, powered by machine learning, generate insights from vast amounts of data, offering predictions for various aspects of our lives from projected possibilities of health disorder \cite{rana2024ai,tutun2023ai} to where criminal activities might occur tonight \cite{shah2021crime} (of course, with much criticism and debate on its ethical implications \cite{brayne2021technologies,ziosi2024evidence}). However, due to their complexity, even experts struggle to fully understand how these models arrive at specific conclusions, connecting to the highly popularized notion of ‘explainability’ in the field of AI \cite{dwivedi2023explainable,rudin2019stop}.

Similarly, fortunetelling employs various methods that are ‘scientifically’ unexplainable—such as connecting to transcendent beings, analyzing birth dates, or reading faces—to interpret personal traits and predict future destiny. In such contexts, despite their differing epistemological foundations, both AI and fortunetelling ultimately serve a common function: helping individuals make sense of the uncertainty.

\textbf{AI Fortune-teller} is a speculative design project inspired by this observation. Presented in video form, we document an experiment in deception that juxtaposes AI and fortunetelling. Participants engaged with what they believed to be a career counseling AI chatbot agent, unaware that the responses were actually crafted by a \textit{mudang} (Korean traditional shaman). Upon learning that the advice came not from an AI but from a \textit{mudang}, all participants maintained their initial stance toward the guidance they received, asserting that, regardless of the source, advice remains just advice—it is ultimately up to them to claim their future and take responsibility for their choices. 

Through these remarks, we aimed to draw the audience’s attention to the notion of human agency in the age of AI. In a world where algorithms shape nearly every aspect of our behavior—from what we purchase on Amazon to what we watch on YouTube and like on Instagram—we sought to highlight the enduring role of human agency. Despite the pervasive influence of technology, individuals still retain the power to make their own decisions. As such, our work invites viewers to reflect on how we engage with technology, how we trust or dismiss its recommendations, and ultimately, how we define our own futures.

\begin{quote}
\textit{``It’s bringing something outside of the human realm. We may admire them, we may trust them. But we just can’t explain how it works!''}
\end{quote}

\section{Creating AI Fortune-Teller}
\subsection{Recruitment}
We recruited seven participants in their 20s and 30s who were navigating career-related concerns. In recruiting participants, the project was presented as a documentary film exploring the usability and perception of AI-based career counseling systems. The recruitment form requested the following information from participants: name, a photograph for camera testing, birthdate, and specific career-related questions they wished to ask. This information, in reality, was requested by the \textit{mudang} to read their characteristics and future. 

All participants for this project were recruited from our acquaintances to minimize the potential shock caused by deception. For instance, culturally, consulting a \textit{mudang} might conflict with the religious beliefs of devout Christian individuals. However, directly inquiring about participants’ religious beliefs could have revealed that the project was not a standard AI-based career counseling test run. To avoid this, only individuals known to have no reservations about consulting a \textit{mudang} were selected.

\subsection{Preparing the Deception}

In preparation for the deception, we undertook two key procedures: (1) preparing the chatbot responses and (2) making the participants believe that those responses were coming from the AI’s analysis of their own data. 

First, we crafted responses that the fake chatbot could send out based on the \textit{mudang}’s readings of the participants. We compiled a one-page profile for each participant using the key information they provided—name, birthdate, photo, and questions they wanted to ask on their career path—and shared it with the \textit{mudang}. 

\begin{figure}
    \centering
    \includegraphics[width=\linewidth]{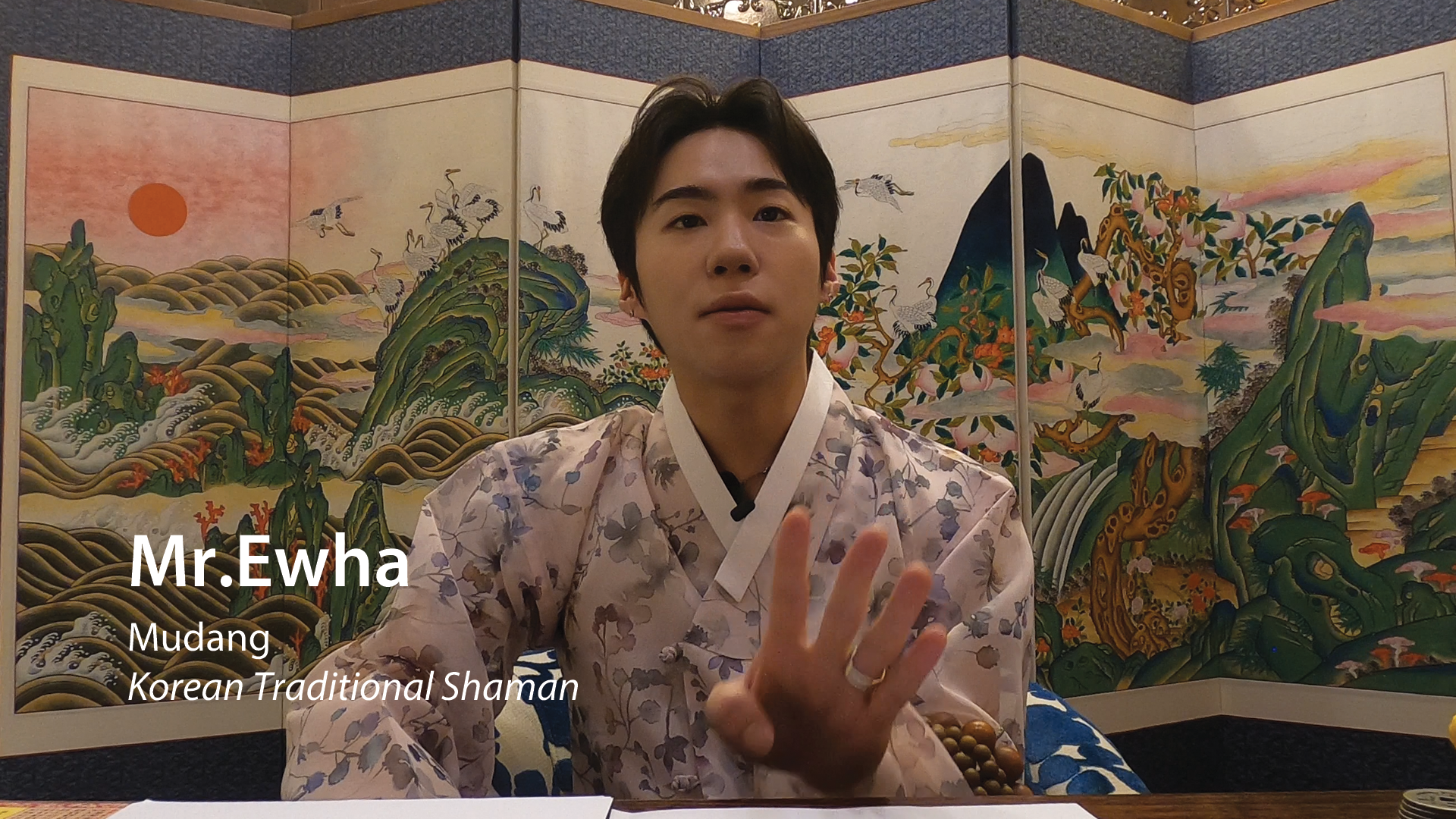}
    \caption{Mr. Ewha, a \textit{mudang} who read the future of our participants}
    \label{fig:mudang}
\end{figure}

Drawing from this, the \textit{mudang} provided detailed readings on their personalities, talents, interests, potential future paths, and answers to their questions. We then transcribed his readings and reorganized them into responses that answered participants’ pre-submitted inquiries and general career-related questions. To ensure a consistent tone that resembled an AI-generated response, we refined the text using ChatGPT.

For instance, if this is what the \textit{mudang} said:

\begin{quote}

\textit{``Is she currently studying? It seems like she’s not a studying-type of person. She’s not the kind of person who would sit still and do desk work. She needs to go out and start a business or do something like that. But she doesn’t have the business sense to succeed alone. She has to meet the right people.''}

\end{quote}

This is what the prepared response looked like once we reworked it:

\begin{quote}

\textit{``Your personality may not be well-suited for routine tasks. If you are currently in a role that requires studying or strict attendance, you may find it unsatisfying. You tend to feel more fulfilled when working independently, such as starting a business. However, the results indicate that running a business may present challenges due to potential gaps in business acumen or managerial skills. Therefore, it will be crucial for you to find reliable partners to collaborate with.''}
    
\end{quote}

Secondly, recognizing that a real career-counseling AI would require some form of data to generate analyses and predictions, we asked participants to complete a 33-item questionnaire resembling a career-related self-assessment test before the session: 

\begin{itemize}
    \item I feel excited and joyful when experiencing something new.
    \item I want to be a person who has both wealth and fame.
    \item I want to maintain a clear work-life balance.
    \item I hope that the work I do makes a positive contribution to society.
    \item I want to become the best in my field.
\end{itemize}

Of course, these responses were not utilized in any way. Their role was to merely deceive our participants.

\subsection{Interacting with the Agent}

Having completed the necessary preparations for participant engagement, the next step was to conduct the actual sessions. Each session was held individually and consisted of a hands-on experience with the chatbot agent, along with pre- and post-usage interviews. 

The session began with questions about participants’ prior experiences with AI-related agents, their understanding of how these systems operate, and the level of trust they place in them. This approach allowed us to explore participants’ expectations and attitudes toward AI agents while also assessing their current understanding of how these technologies function.

Following the interview, the conversation with the AI agent took place. Participants were informed that our system was integrated with Facebook’s Messenger. The participants were asked to send a greeting message (“Hello”) to the AI agent’s Messenger profile to initiate the system.

Once the message was sent, an operator in a separate room responded with a scripted message:

\begin{quote}
\textit{``Hello [Name], I am a career counseling AI. Do you have any questions? You are welcome to ask any questions, including the ones you asked during the recruitment phase.''}    
\end{quote}

Immediately after, the participant’s pre-submitted career-related questions were sent as a list, encouraging them to engage with the responses we had prepared. If a participant asked a question that did not have a pre-scripted response, the operator quickly generated an answer based on the \textit{mudang}’s reading, processed it through ChatGPT for consistency, and sent it. If a response could not be formulated, the participant received a message stating, 

\begin{quote}
\textit{``Information is not sufficient enough to provide an answer.''}    
\end{quote}

This approach maintained the illusion of real-time interaction while ensuring that all responses remained aligned with the \textit{mudang}’s insights.

\begin{figure*}
    \centering
    \includegraphics[width=\linewidth]{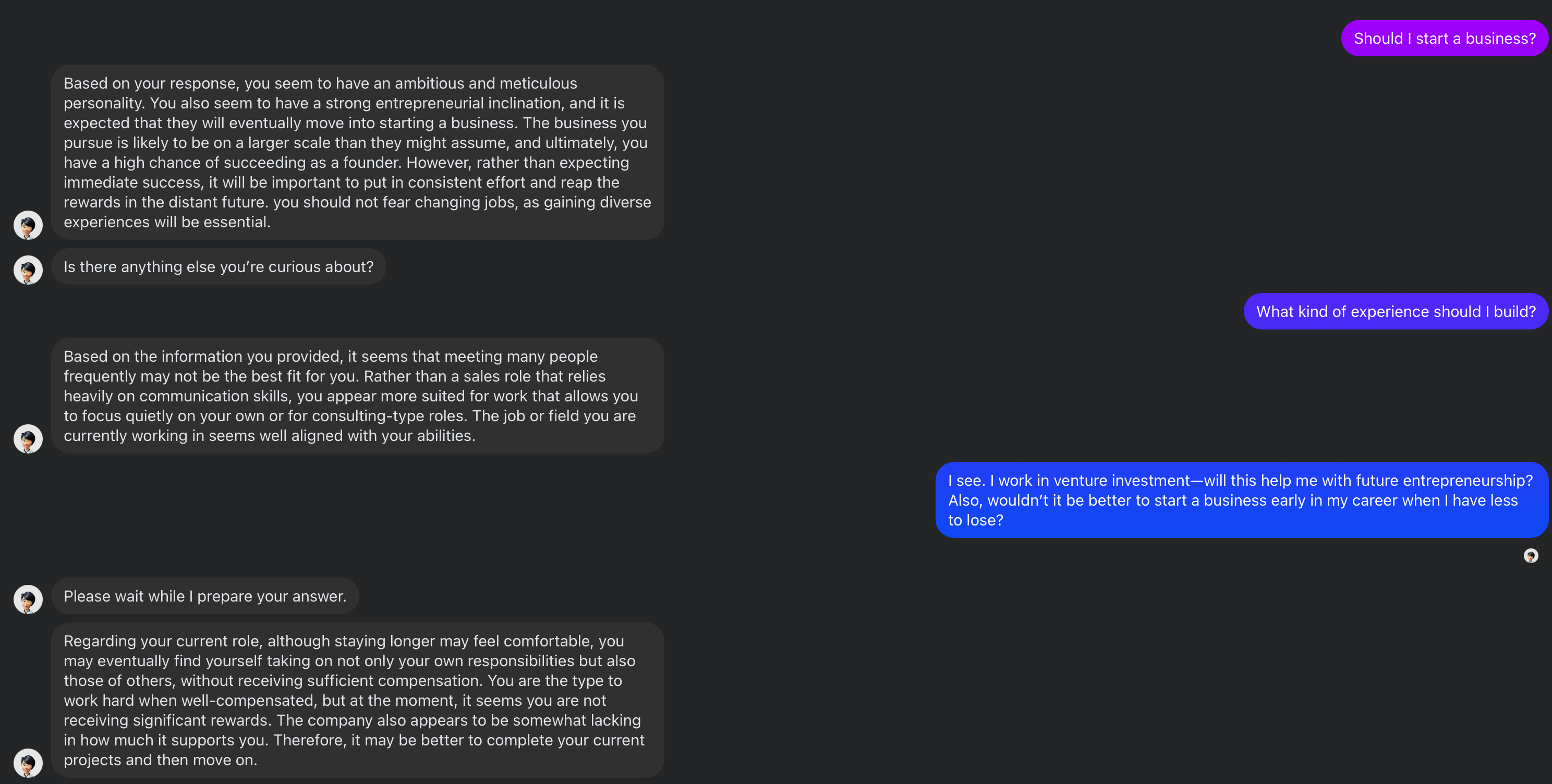}
    \caption{A translated conversation between one of the participants and our ``AI'' agent}
    
    \label{fig:conversation}
\end{figure*}

After several exchanges with the AI agent, we proceeded to ask participants about their experience. We inquired whether they found the responses useful or relevant to their situations and whether the AI agent’s answers influenced their career considerations.

We then asked participants to speculate on the inner workings of the chatbot agent. After their speculation, we finally revealed that the agent was actually giving advice based on a \textit{mudang}’s reading and not the AI. Following the reveal, we inquired about their past experiences with fortunetelling and their expectations of such pre-scientific decision-support methods. 

Through these steps, we sought to further explore the parallels between people’s attitudes toward AI and fortunetelling—how both highly technological agents and various forms of fortunetelling are often used for self-analysis and prediction, how users often are unaware of the inner workings behind such analysis and predictions, and how they try to strike a balance between following such advice and their critical thinking to uncover their true desires.

To conclude the session, we revisited the career advice previously provided by the AI agent and asked whether participants’ perceptions of the guidance (or any decisions they had considered based on it) had changed after learning that the responses originated from a \textit{mudang} rather than an AI. In the end, all participants maintained their initial stance toward the advice. In other words, participants suggested that the nature of the advising agent—whether technical or superstitious—did not ultimately matter; they emphasized that \textit{they themselves} were the ones making the final decisions.

\section{Discussions}
In this section, we step back and reflect on the project from our perspective as creators of this piece. In doing so, we interweave our reflections with resonant excerpts from participants whose words helped shape and challenge our thinking.

\subsection{Seeking Solace in the Face of Uncertainty}
Why do we seek AI or fortunetellers? One possible explanation is that we seek comfort from entities that appear to transcend human reasoning—whether through advanced technology or mystical guidance. Often, this search is driven by the fear of making decisions alone and the desire for reassurance:

\begin{quote}
    \textit{``I think we find them to get some assurance.''}
\end{quote}

\begin{quote}
    \textit{``Doing something because someone told you to do so takes off the burden. You can blame it on them!''}
\end{quote}

Many participants expressed that interacting with such agents, being provided with practical advice and an emotional boost, made them feel supported in their decision-making process. The mere act of having someone—or something—engage with their concerns was often enough to offer a sense of encouragement and reassurance:

\begin{quote}
\textit{``You just go there to hear good things. It’s the power of rooting.''}
\end{quote}

\begin{quote}
\textit{``It felt like it was rooting for me. It gave me courage.''}
\end{quote}

\subsection{The Answer Within Myself}
One of the most significant realizations among participants after the session was that, in many cases, the answer was already within themselves. One way this unfolds is through verbalizing concerns. Speaking about their worries—whether to an AI agent or a \textit{mudang}—allowed participants to recognize their own thoughts and feelings more clearly. Many reflected that simply articulating their concerns to an external entity helped them structure their thoughts and solidify their understanding of what they truly wanted.

\begin{quote}

\textit{``I already had an answer in me. So, I don’t think I was seeking an answer from them. It was more like, I organized the answers that I wanted to hear through asking.''}
    
\end{quote}

Another way this occurred was through reacting to the given responses. Often, receiving an answer (whether from AI or a \textit{mudang}) serves as a prompt rather than a directive. It is not about blindly following advice but about using it to refine one’s own stance. Perhaps we engage in this kind of interaction more often than we realize. When a friend asks whether you want Korean or Mexican food for dinner, you may find yourself unable to choose. Sensing your indecision, your friend declares, “Let’s do Mexican.” And in that instant, you realize that what you truly wanted all along was Korean.

In conversations like this, an external suggestion forces introspection, compelling the person to realize what they truly want. Similarly, some participants shared that when the agent told them not to pursue something they deeply desired, it only motivated them to explore the reasoning further and push forward with greater determination.

These suggest that the efficacy of AI or fortunetellers does not lie in the accuracy of the responses they provide, but rather in the process of interaction—a process that helps individuals clarify their true desires.

\subsection{Making a Decision}
To summarize the above two points, in the face of uncertainty, we seek comfort in guidance from entities beyond human reasoning. Yet, the final decision always falls back to us—whether it is trusting the numbers behind AI’s advice, putting faith in a \textit{mudang}’s predictions, or following our hearts. Ultimately, \textit{we} take ownership of that decision, along with the responsibility that comes with it.

We find ourselves in an era where we are both driven by the pursuit of advanced AI systems like Artificial General Intelligence or strong AI and gripped by fears about their ethical implications and the future they may bring (“The machines will take over!”). Against this backdrop, our work sought to capture the messy, complex ways in which people negotiate, challenge, and ultimately assert their autonomy when interacting with AI, juxtaposing it with something that has long existed to support human decision-making: the pre-scientific practices of fortunetelling and spiritual guidance. 

Of course, our goal is not to directly compare or advocate for one over the other. After all, we cannot deny the efficacy of using predictive algorithms for purposes such as medical predictions. Rather, at the heart of our artwork lies a reaffirmation of everyday human agency in the age of AI, showcasing how our participants (and, by extension, humanity) ultimately make their own choices. Our piece underscores the sacredness of human decision-making, especially in an age filled with both excitement and apprehension about the future AI may bring. In the midst of this uncertainty, we emphasize the importance of striving for an authentic life; one where we continue to claim our agency, shape our own paths, and navigate the unknown on our own terms.

\section{Conclusion: Celebrating uncertainty as a stage for human agency in the age of AI}

After watching the final video, the \textit{mudang} who gave readings to our participants' future expressed deep appreciation and respect for all participants. He shared that the role of a \textit{mudang} has always been to help people live courageously, meaningfully, and as themselves—to guide them in embracing their own paths. Seeing participants engage with his readings in such a way, he felt that the message had been truly understood.

We often turn to AI to make sense of our present and gain clarity in navigating an uncertain future. Yet, no matter how advanced AI becomes, uncertainty will never fully disappear, especially in moments where we have to make deeply subjective choices. In these moments, once again, it is we who make the decision. \textit{We} are the ones who hold the agency, the responsibility, and the courage to take that final step. As our participants beautifully put it:

\begin{quote}
\textit{``If AI tells me not to go there, but if I really, really want to go… I still will. Because this is my dream. Even if I fail, that’s on me. AI could be wrong.''}    
\end{quote}

Through this speculative piece, we wanted to move our conversations on AI beyond the pursuit of efficiency, productivity, and accuracy, deeply resonating with what the field of HCI has already been doing well \cite{suchman1987plans,bodker2006second}. We sought to invite audiences to reflect on what it might mean to design AI that acknowledges and celebrates human agency in an uncertain world. With that, we want to conclude this piece with an empowering statement: no matter how advanced technology becomes, no matter how much AI shapes the world around us, we will continue to navigate our uncertain futures in the most human way imaginable—wonderfully messy and uncertain.

\begin{acks}

We extend our gratitude to everyone who contributed to the creation of this artwork. The artwork was edited by Gyutae Kim and features Ewha Doryeong (our \textit{mudang}), Sungho Chang, Haein Cho, Jin Ho Hwang, Dohyeon Kim, Jungyoon Kim, Hyo Bin Lee, and Ji Won Lee. The locations for the artwork were provided by Default Studio and GT Entertainment. We also thank the Prix Ars Electronica for recognizing the value of our work.

\end{acks}

\bibliographystyle{ACM-Reference-Format}
\bibliography{ref}


\end{document}